# Optical Network Programmability – Requirements and Applications

Achim Autenrieth[(1)], Jörg Peter Elbers[(1)], Thomas Szyrkowiec[(1)(2)], Pawel Kaczmarek[(3)], Wolfgang Kellerer[(2)]

[(1)] ADVA Optical Networking, Martinsried/Munich, Germany
[(2)] Technische Universität München, Munich, Germany
[(3)] ADVA Optical Networking, Gdynia, Poland
Email: aautenrieth@advaoptical.com

*Abstract*—Datacenter operators and internet content providers require optical network programmability to efficiently interconnect distributed datacenters. This paper describes the requirements, applications and use cases for optical network programmability. Based on application scenarios, open northbound APIs with different levels of control and abstraction to address different network operator requirements which are defined. For their illustration, use cases for optical network programmability show current research directions.

*Keywords—component; formatting; style; styling; insert (key words)*

I. INTRODUCTION

In addition to mobile services, cloud services are becoming a main traffic growth factor, and distributed mega datacenters are increasingly requiring flexible optical networks with on-demand self-service for their interconnection [1][2]. Hence, a more programmatic mode of operation and simpler integration into existing datacenter control frameworks is needed. Software-Defined Networking (SDN) is a promising approach to address such programmability. The basic idea of SDN is to separate the control plane from the forwarding plane. The control of multiple forwarding devices is handed over to a logically centralized entity — the SDN controller. This supervisory role gives the controller a holistic view of the network which is an integral requirement for making the network as a whole programmable and allowing for better overall forwarding decisions. By exposing an application programming interface (API), network elements are no longer closed and vendor dependent, but become rather open and programmable. For transport SDN, the control protocol is turning from OpenFlow, which was primarily designed for connection-less, packet-switched L2 and L3 networks, to model-driven APIs based on YANG [3] models. This approach has the advantage that different APIs, like Command-Line Interface (CLI), REST, NETCONF [4], or RESTCONF [5], can be programmatically generated directly from the YANG model. A system vendor can open its proprietary YANG models or additionally support generic, vendor-independent models. The model-driven approach allows to be consistent with different models for different applications & abstraction levels.

In this paper we first define requirements for programmable optical networks, then we describe different network application scenarios, and finally we present ongoing research projects and current results aiming to solve the technical challenges of programmable optical networks.

II. REQUIREMENTS

In this section a number of key requirements for programmable networks are discussed.

The first requirement is **Ease of Use**. Datacenters use state-of-the-art operation systems. Server loads are monitored and optimized, with loads continually being balanced across all available resources within their walls. Traditional network management software is not compatible with datacenter software, thus preventing the latter from truly optimizing end-to-end multilayer packet flows between interconnected datacenters. The network should support a simple programming interface for the automation of repetitive actions based on templates.

**Openness** is an important requirement to avoid hard vendor lock-in. Openness can be achieved at multiple levels. First of all, the network should support multiple SDN controllers, ideally open-source controllers like OpenDaylight. Secondly, by using a model-driven approach, a datacenter operator can use the model and the procedure calls defined in the model to integrate the network into their own operation systems.

**Programmability** of the network is required to support application-driven automation of operational procedures rather than manual configuration and management. To support programmability, application programming interfaces can be generated from the YANG models. The models include service templates and remote procedure calls e.g. for topology dissemination, path calculation, and service provisioning.

Finally, **abstraction** is required for applications like multilayer interworking, multi-vendor interoperability, or network virtualization (delegation of control of a virtual slice of the network to a client).

III. APPLICATIONS FOR PROGRAMMABLE OPTICAL NETWORKS

The level of control and abstraction required for Data Center Interconnect (DCI) customers varies widely. Fig. 1 shows different application scenarios for programmable optical networks for datacenter interconnect.

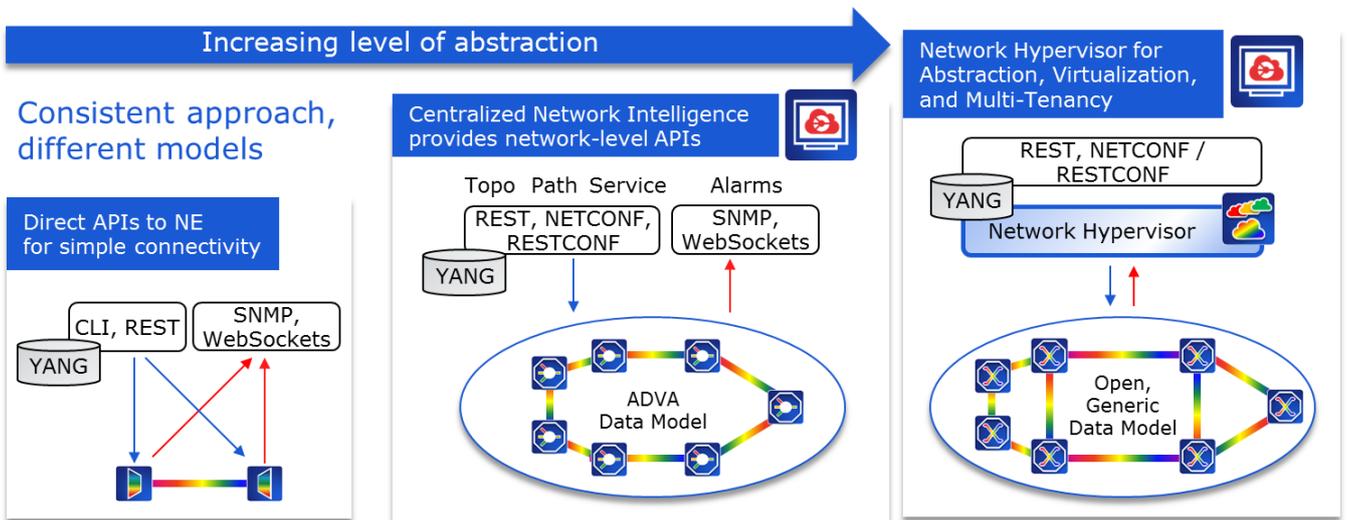

Fig. 1 Application scenarios and corresponding APIs.

*A. Direct APIs to NEs for Simple Connectivity*

For simple connectivity of point-to-point networks, as shown on the left side of Fig. 1, often direct APIs are the right choice. In this case, network elements are provisioned using Command Line Interfaces (CLI), sometimes with scripting routines. Basic Representational State Transfer (REST) interfaces can be used here, as well. The model is typically a vendor-specific or generic network element API, as in these simple application scenarios a low level of abstraction is required.

*B. Network-level APIs for Meshed Networks*

For more complex networks beyond simple point-to-point links, e.g. metro/regional infrastructure networks as shown in the middle part of Fig. 1, APIs with network-level programmability are required. The aforementioned RESTful interface is one example. Another option for the communication with the network elements is the Network Configuration Protocol (NETCONF). NETCONF uses YANG models to represent the configuration and is already popular with routers and switches. RESTCONF attempts to combine the advantages of both REST and NETCONF, by using a RESTful interface running over HTTP for accessing data defined by a YANG mode and stored in a NETCONF data store. These interfaces are provided by a centralized network intelligence.

*C. Network Hypervisor for Abstraction, Virtualization and Multitenancy*

Finally, for more complex applications like multidomain orchestration, or IP over DWDM multilayer interworking, a higher level of abstraction is required. Additionally, virtualization and multitenancy are required if control over a virtual slice of the network should be delegated to a client. A network hypervisor, which acts as an abstraction and virtualization layer as shown in the right part of Fig. 1, supplements the centralized intelligence

The network hypervisor is not a controller in itself, but rather a mediator that lets anyone or anything query and control the transport network. For example, an orchestration application attempting to load balance across virtual machines at different geographically dispersed locations can include the transport network in its optimization.

Another example of abstracting the transport network with a network hypervisor is multitenancy. A typical Service Provider will use a common transport network to serve multiple clients. Currently, each of these clients may have bandwidth connectivity, but they must go through the Service Provider for all network management tasks. A multitenancy capable network hypervisor allows each client to have his own virtual transport network, based upon the real physical assets available. Each client sees his own and only his own portion of the network and can configure it within pre-determined configuration rules. The Service Provider maintains visibility of the entire network. This facilitates a Service Provider to offer control over a slice of the shared network to customers who would have demand for their own manageable network.

IV. USE CASES

In this section two main use cases are discussed based on ongoing research in current funded research projects.

*A. Multidomain orchestration (STRAUSS)*

An important use case for SDN-controlled optical networks is multi-domain orchestrations. The EU/Japan co-funded research project STRAUSS [6] investigates the orchestration of end-to-end Ethernet services over multiple, heterogeneous network domains. Each network domain can use a different transmission technology, like optical packet switching (OPS), or flexi-grid or fixed grid optical circuit switching (OCS), and different control plane protocols such as GMPLS/PCE, or OpenFlow.

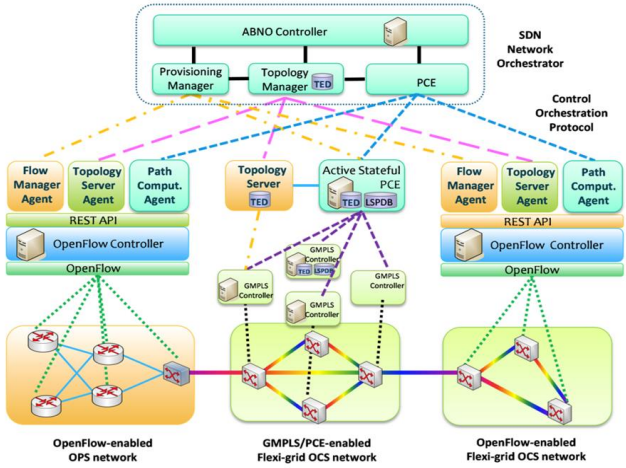

Fig. 2 STRAUSS Multidomain Orchestration

The SDN network orchestrator is communicating with the different domains via the Control Orchestration Protocol (COP). COP has been defined using the YANG modeling language and can be transported using REST or RESTCONF. STRAUSS has open sourced the COP yang models and tools to support COP's adoption in the academic as well as the industry community [7]. COP includes following models:

- Call
- Topology
- Path Computation

### B. Multilayer interworking (ACINO)

ACINO [8] is focusing more on the communication between the application and the orchestrator – including the underlying network controllers. Following an application-centric approach to transport networks it is expected that application requirements can be fulfilled more accurately by propagating individual requirements to the transport layer. Therefore, the offered service is tailored to the application's needs.

The general idea is to create an orchestrator which exposes a set of high-level primitives to the applications interfacing northbound. Through this interface the application can express its intent and receive the needed service. The application's requirements are translated into commands that are directed toward the IP and optical controllers respectively. This utilizes multilayer planning and configuration capabilities of the orchestrator.

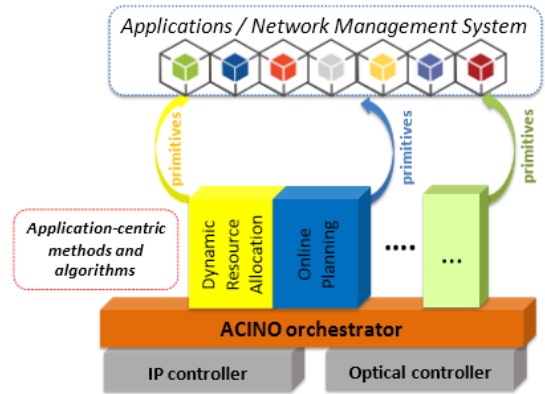

Fig. 3 ACINO Multilayer Interworking

YANG model to communicate abstract TE topologies beween two layers or two domains is defined in the a current internet draft [9].

## V. CONCLUSION

Transport SDN brings programmability & virtualization to optical networks. A programmable optical networking is key to future-proof packet-optical integration and must offer open application programming interfaces, ideally using YANG model-driven approach, to support operation and management systems of network operators as well as open-source SDN controllers. Different levels of control and abstraction allow network programmability from box-level APIs for simple point-to-point connectivity, to network-level APIs for network level services, and finally to abstract APIs for multi-vendor interoperability, multi-domain orchestration, and multilayer interworking.

REFERENCES

[1] Cisco, "Cisco Visual Networking Index: Global Mobile Data Traffic Forecast Update, 2014–2019", 2015
[2] "Cisco Global Cloud Index: Forecast and Methodology 2013–2018 White Paper", 2014
[3] M. Bjorklund, "YANG - A Data Modeling Language for the Network Configuration Protocol (NETCONF)", IETF RFC 6020, Oct 2010
[4] R. Enns, M. Bjorklund, J. Schoenwaelder, and A. Bierman. Network Configuration Protocol (NETCONF). RFC 6421, June 2011.
[5] A. Bierman, M. Bjorklund, K.Watsen, " RESTCONF Protocol", draft-ietf-netconf-restconf-05 (work in progress), IETF Draft, June 2015.
[6] STRAUSS project website, http://www.ict-strauss.eu/
[7] Control Orchestration Protocol (COP), https://github.com/ict-strauss/COP/
[8] ACINO: Application-Centric IP/Optical Network Orchestration, EU H2020 Project, (GA 645127, ICT-2014-1), www.acino.eu
[9] Liu, Bryskin, Beeram, Saad, Shah, "YANG Data Model for TE Topologies", IETF draft, work in progress, draft-ietf-teas-yang-te-topo-00, May 8, 2015
[10] Clemm et al. "A Data Model for Network Topologies",draft-ietf-i2rs-yang-network-topo-01 (work in progress), IETF draft, June 2015

The research leading to these results has received funding from the European Commission within the Seventh Framework Programme (FP7/2007-2013) under grant agreement n.608528, project STRAUSS (www.ict-strauss.eu), and the H2020 Research and Innovation Programme, under grant agreeement n.645127, project ACINO (www.acino.eu).